\documentstyle[12pt]{article}
\newcommand{\be}{\begin{equation}}
\newcommand{\ee}{\end{equation}}
\newcommand{\ba}{\begin{eqnarray}}
\newcommand{\ea}{\end{eqnarray}}

\newcommand{\dsl}
  {\kern.06em\hbox{\raise.15ex\hbox{$/$}\kern-.56em\hbox{$\partial$}}}

\newcommand{\eeq}{\end{equation}}
\newcommand{\ZZ}{{\rm \kern 0.275em Z \kern -0.92em Z}\;}
\begin{document}
\title{Supersymmetric Non-Abelian Born-Infeld Theory}
\author{S.~Gonorazky\thanks{UNLP-FOMEC}\, ,
F.A.~Schaposnik\thanks{Investigador CICBA} \, and
G.~Silva\thanks{CONICET}
\\
{\normalsize\it
Departamento de F\'\i sica, Universidad Nacional de La Plata}\\
{\normalsize\it
C.C. 67, 1900 La Plata, Argentina}}
\date{\hfill}
\maketitle
\begin{abstract}
Using the natural curvature invariants as building blocks
in a superfield construction,
we show  that the use of a symmetric trace is mandatory
if one is to reproduce the square root structure of the
non-Abelian Dirac-Born-Infeld Lagrangian in the bosonic sector.
We also discuss the BPS relations in connection with our supersymmetry
construction.

\end{abstract}


\bigskip

\newpage

Dirac-Born-Infeld (DBI)
type actions arise in the study of low-energy
dynamics of D-branes
\cite{Tse}-\cite{G} (see \cite{Pol}-\cite{Tay}  for a complete
list of references). In the case of superstring theory, one  has to
deal  with a supersymmetric extension of DBI actions and,
when a number of D-branes coincide, there is a symmetry enhancement
\cite{Wi}
and the Abelian DBI action should be generalized to the non-Abelian case.

Several possibilities for extending the Abelian
Born-Infeld action to the
case of a non-Abelian gauge symmetry  have been discussed in the
literature \cite{AN}-\cite{Hashi}. Basically, they differ in the
way the group trace operation is defined. In the string context,
a symmetric trace operation as that  advocated by
Tseytlin \cite{Tse2} seems to be the appropriate one. Among its advantages,
one can mention:

(i) It eliminates unwelcome
odd powers of the curvature, this implying
that the field strength $F$ (although possibly large)
should be slow varying since $ F^3   \sim [D,D]F^2$.
With this kind of Abelian approximation (it
implies commuting $F's$) one can make
contact with the tree level open string effective action.

(ii)
It is the only one leading to an action which is linearized
by BPS conditions and to equations of motion
which coincide with those arising
by imposing the vanishing of the $\beta$-function
for background fields in the open superstring theory \cite{G1}-\cite{B}.

\vspace{0.1 cm}
It
should be mentioned, however,  that there are some unsolved problems concerning
the use of a symmetric trace for the non-Abelian Born-Infeld action.
In particular,    some discrepancies
between the results arising from a
symmetrized non-Abelian Born Infeld theory and the
spectrum to be expected in brane theories are pointed out in ref.\ \cite{HT}.
\vspace{0.19 cm}

As noticed in \cite{B}, the fact that the symmetric trace is
singled out as that leading to  BPS relations should be connected
with the possibility of supersymmetrizing the Born-Infeld theory.
In this respect, we
construct  in this work  the supersymmetric
version of the non-Abelian Dirac-Born-Infeld action and discuss
  the trace issue and   the Bogomol'nyi
structure of the resulting bosonic sector.

Our analysis, close to that
developed in \cite{DP}, extends to the non-Abelian case the results
presented for the  Abelian Supersymmetric Born-Infeld
theory   in ref.\ \cite{GNSS}.

 As it is
well-known, the basic object for constructing supersymmetric
gauge theories is the
(non-Abelian)
gauge vector superfield $V$  which we shall write
(in $d=4$ Minkowski space) in the form
\begin{eqnarray}
& & V(x,\theta,\bar\theta)  =  C + i \theta \chi - i \bar \theta \bar \chi
+ \frac{i}{2} \theta \theta (M + iN) -
\frac{i}{2} \bar \theta \bar \theta (M - iN)
 -\nonumber\\
& &  \theta \sigma^\mu \bar \theta A_\mu +
i \theta \theta \bar \theta (\bar \lambda +\frac{i}{2} \bar{\dsl} \chi)
 -   i \bar \theta \bar \theta \theta (\lambda + \frac{i}{2} \dsl \bar \chi)
+ \frac{1}{2} \theta  \theta  \bar \theta  \bar \theta (D +
\frac{1}{2} \Box C)
\label{1}
\end{eqnarray}
In the case of SUSY Yang-Mills theory,
(generalized) gauge invariance
allows to work  in the Wess-Zumino gauge,
for which $C,\chi,M$ and $N$ are all set
to zero, thus remaining a multiplet with
the gauge field $A_\mu$, the Majorana fermion field $\lambda$ and
the auxiliary real field $D$, all  taking values in the Lie algebra of the
gauge group which we  take for definiteness as $SU(N)$,
\be
A_\mu = A_\mu^a t^a \;\;\; \;\;\;
\lambda = \lambda ^a t^a \;\;\; \;\;\; D = D^a t^a
\label{10}
\ee
with $t^a$ the (hermitian) $SU(N)$ generators,
\be
[t^a,t^b] = i f^{abc} t^c
\label{11}
\ee
\be
{\rm tr} \, t^a t^b =  {\cal N} \delta^{ab}
\label{12}
\ee

It is convenient  to define
a chiral variable $y^\mu$ in the form
\be
y^\mu =x^\mu+i\theta \sigma ^\mu\bar \theta
\label{3}
\ee
so that the usual derivatives $D$ and $\bar D$
can be defined  as
\be
D_\alpha = ~ \frac \partial {\partial \theta ^\alpha }+2i
\left( \sigma ^\mu\bar
\theta \right) _\alpha \frac \partial {\partial y^\mu}, \ \ \ \ \
\bar D_{\dot\alpha} = -\frac{\partial}{\partial \bar\theta^{\dot\alpha}}
\label{4}
\ee
when acting on functions of $(y,\theta,\bar\theta)$ and
\be
D_{\alpha} =  \frac{\partial}{\partial \theta^\alpha},
\ \ \ \
\bar D_{\dot \alpha } = - \frac
\partial {\partial \bar \theta ^{\dot \alpha }%
}-2i\left( \theta \sigma ^\mu\right) _{\dot \alpha }
\frac \partial {\partial y^{\dagger\mu}}
\label{5}
\ee
on functions of $(y^\dagger,\theta,\bar\theta)$.

Generalized
gauge transformation  will be written in the form
\be
\exp(2i\Lambda)= \exp(2i\Lambda^a t^a)
\label{ele}
\ee
where $\Lambda(y,\theta)$ is a chiral left-handed superfield and
$\Lambda^\dagger
(y^\dagger,\bar \theta)$
its right-handed Hermitian conjugate,
\be
\bar D_{\dot \alpha} \Lambda = D_\alpha \Lambda^\dagger = 0
\label{condi}
\ee
Explicitly,
\be
\Lambda(y,\theta) =
\frac{1}{2}(A - iB) + \theta \chi + \theta\theta \frac{1}{2}(F +iG)
\label{ex}
\ee
Here $A, B, F, G$ are real scalar fields and $\chi$ is
a Majorana spinor.
Under such a transformation, superfield
$V$ transforms as
\be
\exp(2V) \to \exp (-2i\Lambda^\dagger) \exp(2V) \exp (2i\Lambda)
\label{12b}
\ee

{}From $V$, the  non-Abelian chiral superfield $W_\alpha$ can be constructed,
\begin{equation}
W_\alpha \left( y,\theta \right) =
\frac{1}{8}
\bar D_{\dot\alpha} \bar D^{\dot\alpha} \exp(-2 V) D_\alpha  \exp(2 V)
\label{2}
\end{equation}
In contrast with (\ref{12b}), under a gauge transformation $W_\alpha$
transforms covariantly,
\be
W_\alpha \to \exp (-2i\Lambda) W_\alpha \exp (2i\Lambda)
\label{12c}
\ee
Concerning the hermitian conjugate, $\bar W_\alpha$, it transforms as
\be
\bar W_{\dot \alpha} \to \exp (-2i\Lambda^\dagger)
\bar W_{\dot\alpha} \exp (2i\Lambda^\dagger)
\label{12d}
\ee
Written in components, $W_\alpha$ reads
\begin{equation}
W_\alpha \left( y,\theta \right) =i\lambda _\alpha
-\theta_\alpha D
-\frac{i}{2}\left( \theta\sigma ^\mu\bar \sigma ^\nu \right) _\alpha
F_{\mu\nu}
-\theta \theta \left( \not\!\nabla \bar
\lambda \right) _\alpha
\label{6}
\end{equation}
with
\be
F_{\mu \nu} = \partial_\mu A_\nu - \partial_\nu A_\mu + i[A_\mu,A_\nu]
\label{8}
\ee
and
\be
\left( \not\!\nabla \bar
\lambda \right) _\alpha = \left(\sigma^\mu\right)_{\alpha \dot \alpha}
\left(
\partial_\mu \bar \lambda^{\dot \alpha} + i [A_\mu,\bar \lambda^{\dot \alpha}]
\right)
\label{nuevi}
\ee
As it is well-known, the SUSY
extension of $N=1$  Yang-Mills theory can
be constructed  from $W$ by considering
$W^2$ and its Hermitian conjugate $\bar W^2$. Indeed,
$W^2$ reads
\begin{eqnarray}
W^\alpha W_\alpha &=& -\lambda \lambda -
i (\theta \lambda D + D \theta\lambda) +  \frac{1}{2} (
\theta \sigma^\mu \bar \sigma^\nu \lambda F_{\mu \nu} +
F_{\mu \nu}\theta \sigma^\mu \bar \sigma^\nu \lambda
) + \nonumber\\
& & \theta \theta\left(- i\lambda \not\! \nabla \bar \lambda
-i(\not\! \nabla \bar \lambda)\lambda +
D^2 - \frac{1}{2}(F_{\mu\nu}F^{\mu \nu} +
i \tilde{F}_{\mu\nu}F^{\mu \nu}) \right)
\label{9}
\end{eqnarray}
with
\be
\tilde F_{\mu \nu} = \frac{1}{2} \varepsilon_{\mu \nu \alpha \beta}
F^{\alpha \beta}
\label{dual}
\ee
Or, writing explicitly the $SU(N)$ generators,
\begin{eqnarray}
& & W^\alpha W_\alpha = \{t^a,t^b\}
\!\left(\! -\frac{1}{2}
\lambda^a \lambda^b
- i\theta \lambda^a D^b
+ \frac{1}{2} \theta \sigma^\mu\bar\sigma^\nu \lambda^aF_{\mu \nu}^b
- \right.\nonumber\\
& & \left. i\theta \theta
 \lambda^a \sigma^\mu(\delta^{bc}\partial_\mu
+ f^{bcd} A_\mu^d) \bar \lambda^c
 +  \frac{1}{2}\theta \theta
\left(D^aD^b - \frac{1}{2}(F_{\mu\nu}^aF^{b\,\mu \nu } +
i \tilde{F}_{\mu\nu}^aF^{b\,\mu \nu})\! \right)\right) \nonumber\\
\label{13}
\end{eqnarray}
where
\be
\{t^a,t^b\} = t^at^b + t^b t^a
\label{14}
\ee

{}From eq.(\ref{9}) and an analogous one for  $\bar W_{\dot\alpha} \bar
W^{\dot\alpha}$
one sees  that the supersymmetric Yang-Mills Lagrangian can be written
in the form
\be
L_{SYM} =  \frac{1}{4e^2 {\cal N}}
{\rm tr} \!\!\int\left(
d^2\theta W^2  +  d^2\bar  \theta \bar W^2
\right)
\label{15rere}
\ee
with an on-shell purely bosonic part giving
\be
\left. L_{SYM}\right \vert_{bos}= -\frac{1}{4e^2} F_{\mu \nu}^a  F^{a \mu \nu }
\label{re15}
\ee

We are ready now  to extend the treatment in \cite{DP}-\cite{GNSS}
and find   a general gauge invariant  non-Abelian $N=1$ supersymmetric
Lagrangian  of the DBI type. This Lagrangian will be basically constructed
in terms of $W$, $\bar W$ and $\exp(\pm 2V)$. It is important to note
that at this stage the trace operation
on internal ($SU(N)$)  indices to be used in order to
define a scalar Lagrangian could differ,  in principle,
from the ordinary trace  ``${\rm tr}$''
defined in (\ref{12}) and used in eq.(\ref{15rere}).

In order  to get a DBI like Lagrangian
(written as space-time determinant)
in the bosonic sector of the theory,  one should
include terms which cannot be  expressed in terms of
$F_{\mu\nu}F^{\mu\nu}$ and
$F_{\mu\nu} \tilde F^{\mu\nu}$ like for example those containing
$F^4 \equiv F_{\mu}^{ \alpha}
F_{\alpha}^{ \beta}F_{\beta}^{ \gamma}F_{\gamma}^{ \mu} $. Indeed,
ignoring for the moment ambiguities arising in the definition of
a non-Abelian space-time determinant,  one has, concerning even powers,
\be
-\left. \det\left(g_{\mu \nu} I + F_{\mu \nu}\right) \right\vert_{even~powers}=
I
+ \frac{1}{2} F^2 +  \frac{1}{4}(\frac{1}{2}(F^2)^2 - F^4)
\label{ddd}
\ee
In the Abelian case, the $F^4$  term in the r.h.s.
of eq.(\ref{ddd}) can be written in terms of $F^2$ and
 $F\tilde F$
but this is not the case in the non-Abelian case. Moreover, odd powers of $F$
which were
absent in the former are present in the latter case.

Let us start at this point our search for a
  supersymmetric extension of the non-Abelian DBI model.
To begin with,  in order to get higher (even) powers of
$F_{\mu \nu} F^{\mu \nu}$ and
$\tilde F_{\mu \nu} F^{\mu \nu}$  which   necessarily
arise in a DBI-like  Lagrangian,  we shall
have to include higher powers of $W$ and $\bar W$
combined in such a form as to respect gauge-invariance. In the Abelian case,
this was
achieved  by combining $W^2 \bar W^2$ with adequate powers of
$D^2W^2$ and $\bar D^2 \bar W^2$
\cite{DP}-\cite{GNSS}. In the present non-Abelian case, in view of
transformation laws (\ref{12b}),(\ref{12c}),(\ref{12d}),   the situation is a
little more involved .  Consider then
the possible gauge-invariant superfields that can give rise to quartic terms.
There are  two natural candidates,
\be
Q_1 =  \!\int\! d^2\theta d^2 \bar \theta W^\alpha W_\alpha \exp(-2V) \bar
W_{\dot \beta}
\bar W^{\dot \beta}  \exp(2V)
\label{q1}
\ee
\be
Q_2 =    \int\! d^2\theta d^2 \bar \theta W^\alpha  \exp(-2V) \bar W^{\dot
\beta}
 \exp(2V)
W_\alpha
\exp(-2V) \bar W_{\dot \beta}
 \exp(2V)
\label{q2}
\ee
with purely bosonic components
\be
{\rm tr}\left.Q_1\right\vert_{bos} =  \frac{1}{4}\left(
{\rm tr}( F^2)^2 + {\rm tr} (F\tilde F)^2
\right)
\label{qq1}
\ee
\be
{\rm tr}\left.Q_2\right\vert_{bos} =  \frac{1}{4}\left(
{\rm tr}F_{\mu\nu} F_{\rho\sigma} F^{\mu\nu} F^{\rho\sigma}
+
 {\rm tr} F_{\mu\nu} F_{\rho\sigma} \tilde F^{\mu\nu} \tilde F^{\rho\sigma}
\right)
\label{qq2}
\ee

One can see now that a particular combination of $Q_1$ and $Q_2$  generates
the quartic terms one expects in the expansion of a square root DBI Lagrangian,
provided this last is defined using a symmetric trace. Indeed,
one has
\be
\left. \frac{1}{24}{\rm tr}\left( 2 Q_1+
Q_2\right) \right \vert_{bos}
=  {\rm Str}
\left.\left(1 - \sqrt{1 + \frac{1}{2} (F_{\mu \nu}, F^{\mu \nu})
-\frac{1}{16} (F_{\mu \nu} , \tilde F^{\mu \nu})^2}\,
\right)\right\vert^{4th~ord.}
\label{sr}
\ee
where
\be
{\rm Str} \left(t_1,t_2,\ldots, t_N
\right) = \frac{1}{N!}\sum_{\pi} {\rm tr} \left( t_{\pi(1)}
 t_{\pi(2)} \ldots t_{\pi(N)}\right) \, ,
\label{tra}
\ee
Another feature in favour of using the symmetric trace is
that it gives
the natural way of
resolving ambiguities arising in the definition of the DBI
Lagrangian   as a determinant.
 Indeed, one has
 \cite{tse},\cite{B},
\be
{\rm Str}
\left(1 - \sqrt{1 + \frac{1}{2} (F_{\mu \nu},F^{\mu \nu})
-\frac{1}{16} (F_{\mu \nu},\tilde F^{\mu \nu})^2}\,
\right) = {\rm Str}\left(1 -  \sqrt{\det(g_{\mu \nu} + F_{\mu \nu})}
\right)
\label{chu}
\ee
with the r.h.s. univoquely defined through the Str prescription.

Eq.(\ref{sr}) is one of the main steps in our derivation:
it shows that
 in order to reproduce
the quartic term in the expansion  of a DBI-type square root, one has to choose
a
particular combinacion of the  {normal trace}. But
this combination of normal traces corresponds precisely to the
{symmetric trace},
originally proposed by Tseytlin \cite{tse} for the DBI theory in order to make
contact
with the low energy effective theory arising from superstring theory.  It is
worthwhile to notice
that   the r.h.s. of (\ref{sr}) vanishes for $F = \pm i\tilde F$. This will
guarantee, at least
at the quartic order we are discussing up to now,
that  the supersymmetric Lagrangian
will reduce to SUSY Yang-Mills when the
Bogomol'nyi bound (in the Euclidean version) is saturated, as it should be
\cite{HT}-\cite{Hashi}, \cite{CS}.

The analysis above was made for the purely bosonic sector.  It is then natural
to
extend it by considering the complete superfield combination $2Q_1 +  Q_2$ with
a
trace that
again accomodates in the form of a symmetric trace
\begin{eqnarray}
\frac{1}{3}
(2\!\!\!\!\!\!\!\!\! & & \!{\rm tr}  Q_1 +  {\rm tr}  Q_2)  = \nonumber\\
& & {\rm Str}  \left(
W^\alpha,W_\alpha, \exp(-2V) \bar W_{\dot \beta}
 \exp(2V), \right.
\left. \exp(-2V) \bar W^{\dot \beta}
\exp(2V)
\right)
\label{sr2}
\end{eqnarray}

Now, in order to construct higher powers of $F^2$ and $F\tilde F$,
necessary to obtain the DBI Lagrangian, we
define,
extending the treatment in  \cite{GNSS},
 superfields $X$ and $Y$,
\begin{eqnarray}
X & = &\frac{1}{16}
\left(
\bar D^2
\left(
\exp(-2V)
\bar W_{\dot \alpha} \bar W^{\dot \alpha}
\exp(2V)
\right) +
\right.\nonumber\\
& &
\left.
 \exp(-2V) D^2 \left(\exp(2V) W^{\alpha} W_{\alpha} \exp(-2V)
\right)
\exp(2V)
\right)
\label{x}
\end{eqnarray}
\begin{eqnarray}
Y &=& \frac{i}{32}
\left(
\bar D^2
\left(\exp(-2V)
\bar W_{\dot \alpha} \bar W^{\dot \alpha} \exp(2V)
\right)
 -  \right.\nonumber\\
& & \exp(-2V)\left.  D^2\left(\exp(2V) W^\alpha W_\alpha \exp(-2V)
\right) \exp(2V)
\right)
\label{y}
\end{eqnarray}
Both fields transform like $W_\alpha$  under generalized gauge transformations
\be
X \to \exp(-2i\Lambda) X \exp(2i\Lambda) ~, ~~~  Y \to \exp(-2i\Lambda) Y
\exp(2i\Lambda)
\label{t}
\ee
and their $\theta = 0$ component give, as in the Abelian case,  the
two basic invariants
\be
X\vert_{\theta =0} = \frac{1}{4} F_{\mu \nu}  F^{\mu \nu} ~, ~ ~ ~
Y\vert_{\theta =0} = \frac{1}{8} \tilde F_{\mu \nu}  F^{\mu \nu}
\label{b}
\ee

Inspired  in
the result (\ref{sr2}) obtained in order to reproduce the adequate quartic
power in $F$
and $\tilde F$,
we propose the following  supersymmetric non-Abelian Lagrangian as a candidate
to reproduce the DBI dynamics in its bosonic sector,
\begin{eqnarray}
& & L_{S}  =  L_{SYM} + \left( \sum_{n,m} C_{nm} \int d^4\theta {\rm Str}
\left(W^\alpha ,
W_\alpha , \right. \right.
\nonumber\\
& & \left. \left. \exp(-2V)\bar W_{\dot \beta} \exp(2V)
,\exp(-2V)\bar W^{\dot \beta}\exp(2V),X^n,Y^m
\right) + {\rm h.c.} \right)
\label{L}
\end{eqnarray}
The arbitrary coefficients $C_{nm}$ remain to be determined.
One should retain at this point  that expression (\ref{L})
gives a general Lagrangian corresponding to the supersymmetric
extension of a  bosonic gauge invariant Lagrangian
depending on the field strength  $F$
through
the algebraic invariants $F_{\mu\nu} F^{\mu \nu}$ and
$\tilde F_{\mu\nu} F^{\mu \nu}$, in certain
combinations constrained by supersymmetry. The Abelian
version of (\ref{L})
was engeneered in \cite{DP}-\cite{GNSS}
so that the Dirac-Born-Infeld
Lagrangian could be reproduced by an appropriate
choice of coefficients $C_{nm}$. The same happens in  the non-Abelian case:
a particular choice of $C_{nm}$
corresponds to the non-Abelian  Born-Infeld theory,
\begin{eqnarray}
 C_{0\,0}  & = & \frac{1}{8}
\nonumber\\
  C_{n-2m \,2m} & = & {(-1)^m} \sum_{j=0}^{m} \left (\matrix{
n+2-j\cr
j\cr
}\right ) q_{n+1-j}
\nonumber\\
  C_{n \,2m+1} & = & 0 ~,
\label{coe}
\end{eqnarray}
\begin{eqnarray}
q_0 &=&  -\frac{1}{2}
\nonumber\\
q_n & = & \frac{(-1)^{n+1}}{4n} \frac{(2n-1)!}{(n+1)!(n-1)!}  ~ ~ ~ {\rm for} ~
{}~ n\geq 1
\label{q}
\end{eqnarray}

With this choice one has for the purely bosonic part of  Lagrangian
(\ref{L}),
\be
\left. L_S\right\vert_{bos}
= {\rm Str} \left( 1 - \sqrt{-\det(g_{\mu \nu} + F_{\mu \nu})}
\right)
\label{LL}
\ee
This is the  non-Abelian Dirac-Born-Infeld Lagrangian
in the form originally proposed in ref.\cite{tse}.

As in the Abelian case, there are other
choices of coefficients $C_{mn}$
 which also give consistent causal supersymmetric gauge
theories with non-poly\-no\-mial gauge-field dynamics. In particular,
the alternative proposal for a $SO(N)$ DBI action recently presented
in \cite{RT} should correspond to one of such choices.

We have then been able to construct a $N=1$ supersymmetric Lagrangian
(eq.(\ref{L})) within the superfields formalism, with a bosonic part expressed
in terms of the square root of $\det (g_{\mu \nu} + F_{\mu \nu})$. We have
employed the natural curvature invariants as building
blocks in the superfield construction arriving to a Lagrangian which,
in its bosonic sector, depends only on the invariants $F_{\mu\nu}F^{\mu \nu}$
and  $F_{\mu\nu}\tilde F^{\mu \nu}$ and can be expressed in terms
of the symmetric trace of a determinant.  Odd powers of the field
strength $F$ were excluded in our treatment due to the fact that it is not
possible to construct a superfield functional of $W$ ($\bar W$) and $D W$
($\bar D \bar W$) containing $F^3$ terms in its higher $\theta$ component.

As mentioned above, the trace structure of the non-Abelian Born-Infeld
theory was fixed in \cite{B}
by demanding  the action to be linearized by
BPS-like configurations (instantons, monopoles, vortices). In the present
work we   have seen
 that the symmetric trace naturally arises in the superfield formalism in
the route to the construction of the square root Dirac-Born-Infeld  Lagrangian.
This
confluence of results
is nothing but the manifestation of the
well-known connection between supersymmetry and BPS relations. Then,
in order to complete our work, we shall now describe the BPS aspects in the
model.

For definiteness we shall concentrate on instanton configurations
in $d=4$ dimensional space-time. In the Wess-Zumino gauge, the
$N=1$ supersymmetry vector multiplet is $(A_\mu,\lambda,D)$, with
$\lambda$ a Majorana fermion. Now, in order to
look for BPS relations, we should consider a $N=2$ supersymmetric model
which includes, appart from these fields, those belonging
to a chiral scalar multiplet.
Indeed, in analogy with what   was done to obtain the $N=1$ general
supersymmetric Lagrangian (\ref{L}), one can construct a general
$N=2$ SUSY Lagrangian by adding to the vector multiplet a chiral
multiplet as in the  $N=2$ SUSY Yang-Mills Lagrangian case.
We shall not detail this construction here but just consider the
relevant $N=2$ SUSY transformation laws in order to derive BPS
relations.

A complete $N=2$ vector multiplet
can be accomodated in terms of the fields described
above  in the form
 $(A_\mu,\lambda,\phi,D,F)$ with
$\lambda$ now a Dirac fermion, $\lambda = (\lambda_1,\lambda_2)$,
  $\phi$ a complex scalar, $\phi = M + iN$,  $D$ and $F$
auxiliary real fields.
The gaugino supersymmetric transformation law reads
(we call  $\xi = (\xi_1,\xi_2)$ the $N=2$
 transformation parameter)
\be
\delta \lambda_i = (\Gamma^{\mu \nu}F_{\mu\nu} + \gamma_5D) \xi_i +
i \varepsilon_{ij}
(F + \gamma^\mu\nabla_\mu (M + \gamma_5N))\xi_j
\label{sus}
\ee
where
\be
\Gamma^{\mu \nu} = \frac{i}{4} [\gamma^\mu,\gamma^\nu]
\label{G}
\ee
Instanton configurations correspond to $D=F= \phi=0$ so
that (\ref{sus}) simplifies
to
\be
\delta \lambda = \Gamma^{\mu \nu}F_{\mu\nu} \xi
\label{be}
\ee
or
\be
\delta \lambda = \frac{1}{2}\Gamma^{\mu \nu}(F_{\mu\nu}
+ i \gamma_5 \tilde{F}_{\mu\nu} )\xi
\label{be2}
\ee
In order to look for BPS relations one imposes as usual  $\delta \lambda = 0$
thus obtaining
\begin{eqnarray}
(F_{\mu\nu} + i\tilde{F}_{\mu\nu})\xi_1 & = & 0 \nonumber\\
(F_{\mu\nu} - i \tilde{F}_{\mu\nu})\xi_2 & = & 0
\label{be3}
\end{eqnarray}
In Euclidean space, eqs.(\ref{be3}) become
\begin{eqnarray}
(F_{\mu\nu} + \tilde{F}_{\mu\nu})\xi_1 & = & 0 \nonumber\\
(F_{\mu\nu} - \tilde{F}_{\mu\nu})\xi_2 & = & 0
\label{be4}
\end{eqnarray}
with $\xi_1$ and $\xi_2$   two Euclidean Weyl fermion independent parameters.
As usual, these conditions lead to instanton or anti-instanton self-dual
equations
\be
F_{\mu\nu} = \pm  \tilde{F}_{\mu\nu}
\label{inst}
\ee
each one of its solutions breaking  half of the supersymmetries.

The fact that Yang-Mills self-dual equations arise also when the
dynamics of the gauge field is governed  by a
non-Abelian Born-Infeld Lagrangian  was already observed in
\cite{G1}-\cite{Hashi}. In the context of supersymmetry, this can be
understood following \cite{CS} where it is shown how the
supersymmetry transformation law for the gaugino
(and for the Higgsino in
the case of the example discussed in \cite{CS}), together with the
(algebraic) equation of motion for the auxiliary fields, make the
BPS relations remain unchanged irrespectively of the specific choice
for the gauge field Lagrangian. Moreover, one can see that
the $N=2$ SUSY charges
for a general non-polynomial theory, obtained via the Noether construction,
 coincide, {\it on shell}, with those  arising in Maxwell or
Yang-Mills theories.

In summary, using the superfield formalism, we have derived a
supersymmetric non-Abelian Dirac-Born-Infeld Lagrangian which
shows the expected  BPS structure, namely that of the (normal)
Yang-Mills theory.  In our construction, we have seen that the natural
superfield
functionals from which supersymmetric non-Abelian gauge theories
are usually built,  combine in the adequate, square root DBI form
in such a way that the symmetric trace is singled out as the one
to use in defining a scalar superfield Lagrangian.
  It should be stressed that
the fact that the purely bosonic Lagrangian depends on the
basic invariants $F^2$ and $F\tilde F$ and not on odd powers of $F$
is not the result of the choice of a symmetric trace
but  the consequence of using $W$ and $DW$ as
building blocks for the supersymmetric Lagrangian.
Finally, let us mention
that not only
the supersymmetric DBI Lagrangian but a whole family of non-polynomial
Lagrangians are then included in our main result, eq.(\ref{L}) and
all of them are linearised by BPS configurations which coincide with
those of the normal Yang-Mills theory.

{}~

{}~

\underline{Acknowledgments}:
 G.S. would like to thank Dominic  Brecher
for helpful e-mail correspondence. We all would like to thank
Adri\'an Lugo and Carlos N\'u\~nez for helpful comments and
discussions. This work is
partially  supported by CICBA, CONICET (PIP 4330/96) , ANPCyT
(PICT 97 No:03-00000-02285),
Fundaci\'on Antorchas,   Argentina and a Commission of the European Communities
contract No:C11*-CT93-0315.


\begin{thebibliography}{99}
\bibitem{Tse} E.S.~Fradkin and A.A.~Tsey\-tlin, Phys. Lett. {\bf 163B} (1985)
123.
\bibitem{tse} A.A. Tseytlin, Nucl. Phys. {\bf B276} (1986) 391.
\bibitem{Lei} R.G.~Leigh, Mod. Phys. Lett. {\bf A4} (1989) 2767 and 2073.
\bibitem{CM} C.G.~Callan and J.M.~Maldacena, Nucl. Phys. {\bf B 513} (1998)
198.
\bibitem{G} G.~Gibbons, Nucl. Phys. {\bf B 514} (1998) 603.
\bibitem{Pol} J.~Polchinski,  in {\it Fields, Strings and Duality}
(TASI 96) p.293, eds.
C.~Efthimiou and B.~Greene
World. Sci. 1997.
\bibitem{Tay} W.~Taylor IV, {\it Lectures on D-Branes, Gauge Theory and
M(atrices)},
hep-th/9801182.
\bibitem{Wi} E.~Witten, Nucl. Phys. {\bf B460} (1996) 335.
\bibitem{AN} A.~Abouelsaood, C.G.~Callan, C.R.~Nappi and S.A.~Yost,
Nucl. Phys. {\bf B280} (1987) 599.
\bibitem{H} T.~Hagiwara, Nucl. Phys. {\bf B 189} (1981) 135.
\bibitem{NS1} K.~Shiraishi and S.~Hirenzaki,
Int. Jour. of Mod. Phys. {\bf A6}
(1991) 2635.
\bibitem{Tse2} A.A.~Tseytlin, Nucl. Phys. {\bf B501} (1997) 41.
\bibitem{G1} J.P.~Gauntlett, J.~Gomis and P.K.~Townsend, JHEP {\bf 01}
(1998) 003.
\bibitem{BP} D.~Brecher and M.J.~Perry, Nucl. Phys. {\bf B527} (1998) 121.
\bibitem{B}  D.~Brecher, {\it BPS states of the Non-Abelian
Born-Infeld action}, hep-th/9804180.
\bibitem{HT} A.~Hashimoto and W.~Taylor IV, Nucl. Phys. {\bf 503} (1997)
193.
\bibitem{Hashi} A.~Hashimoto, Phys. Rev. {\bf D57} (1998) 6420.
\bibitem{DP}  S.~Deser and R.~Puzalowski, {\it J. Phys. }{\bf A13 }
(1980) 2501.
\bibitem{GNSS} S.~Gonorazky, C.~N\'u\~nez, F.A.~Schaposnik
and G.~Silva, Nucl.Phys. {\bf B531}
(1998) 168.
\bibitem{CS} H.~Christiansen, C.~N\'u\~nez and F.A.~Schaposnik,
{ \it Uniqueness of Bogomol'nyi equations and Born-Infeld
like Supersymmetric theories},
hep-th/9807197, Phys. Lett {\bf B} in press.
\bibitem{RT} M.~Rocek and A.A.~Tseytlin, {\it Partial
breaking of global D=4 supresymmetry, constrained
superfields, and 3-brane actions}, hep-th/9811232.
%
\end{thebibliography}
\end{document}